\documentclass[conference]{IEEEtran}
\usepackage{graphicx}

\begin{document}

\title{An Architecture for Embedded Systems Supporting Assisted Living}

\author{\IEEEauthorblockN{Daniela Micucci}
\IEEEauthorblockA{Department of Informatics, Systems and Communication\\
University of Milano Bicocca\\
Email: daniela.micucci@disco.unimib.it}
\and
\IEEEauthorblockN{Marco Mobilio}
\IEEEauthorblockA{Department of Informatics, Systems and Communication\\
University of Milano Bicocca\\
Email: marco.mobilio@disco.unimib.it}
}

\maketitle

\begin{abstract}
The rise in life expectancy is one of the great achievements of the twentieth century. 
This phenomenon originates a still increasing interest in Ambient Assisted Living (AAL) technological solutions that may support people in their daily routines allowing an independent and safe lifestyle as long as possible.  AAL systems generally acquire data from the field and reason on them and the context to accomplish their tasks. Very often, AAL systems are vertical solutions, thus making hard their reuse and adaptation to different domains with respect to the ones for which they have been developed. In this paper we propose an architectural solution that allows the acquisition level of an ALL system to be easily built, configured, and extended without affecting the reasoning level of the system. We experienced our proposal in a fall detection system.
\end{abstract}


%
\IEEEpeerreviewmaketitle

\section{Introduction} \label{sec:intro}
In coming decades, population is set to become slightly smaller in more developed countries, but much older. 
This phenomenon, known as \emph{population ageing}~\cite{un_world_2013}, has generated a growing interest in Ambient Assisted Living (AAL)~\cite{kleinberger_ambient_2007}, which encompasses technological solutions supporting elderly people in their daily life at their homes. The structure of an AAL system generally includes a level in charge of acquiring and making available data from the field, and a level in charge of realizing the application logic. 
Usually, AAL system are implemented as vertical solutions in which there is not a clear separation between the two main levels~\cite{calvaresi_exploring_2016}. This rises several issues, among which a scarce reuse of the system components since their responsibilities overlap, and a scarce liability to software evolution mostly because data is strongly coupled with its source. 
To promote reusability and evolution, an AAL system should keep accurately separated issues related to acquisition from those related to reasoning. It follows that data, once acquired, should be completely decoupled from its source. This allows to change the physical characteristics of the sources of information without affecting the application logic level. Moreover, the basic acquisition mechanisms (triggering sources at specified frequencies, and distributing the acquired data) should be kept separated from the software that interacts with the specific source (i.e., the software driver). This allows to reuse the basic mechanisms and to program the drivers for the needed sensors only. If a new or different sensor is required, it suffices to add/change the sensor driver and to properly configure the basic mechanisms so that the change can be implemented.

%
In this paper we propose TANA (Timed Acquisition and Normalisation Architecture), an architecture for the design of the acquisition level that overcomes the limitation of traditional solutions. The architecture includes two layers: the \emph{physical acquisition layer} in charge of easily configuring and managing heterogeneous sensors, and the \emph{normalization layer} in charge of decoupling sources from samples so to present the application logic layer with a unified representation of the acquired data. 


Both the layers have been developed and tested in a real scenario concerning the detection of falls.

\section{The TANA Architecture}
The main goal of TANA is to enable applications to reason on domain specific issues disregarding information about the physical nature and the positioning of the sources of data. 

\begin{figure}[h]
\centering
\includegraphics[scale=0.4]{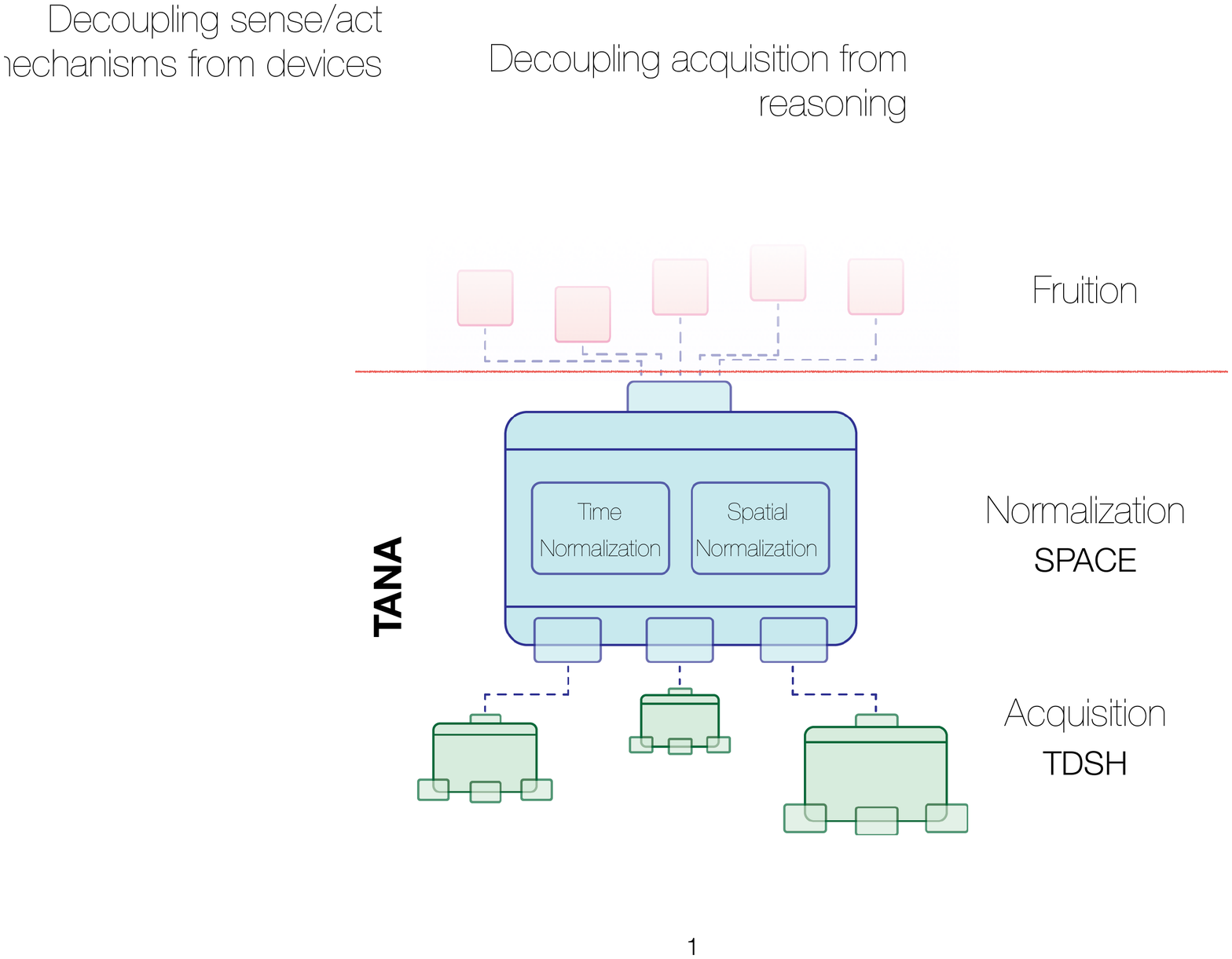}
\caption{The TANA architecture}
\label{fig:TANA}
\end{figure}

Figure~\ref{fig:TANA} shows the proposed architecture: the \emph{physical acquisition layer} is in charge of interfacing with the sensors, the \emph{normalization layer} is in charge of making the acquisitions in an homogeneous format and of decoupling them from their sources. The \emph{fruition layer} is out of the scope of the architecture and includes the AAL applications that use the data from the normalization layer in order to perform application logic reasoning. 

The architecture reifies the two layers by means of the following sets of architectural abstractions: \textit{Time Driven Sensor Hub} (TDSH), which includes the elements for developing timed acquisition systems that are easily configurable for what concerns both the type of the sensors needed and their acquisition frequencies; \textit{Subjective sPaces Architecture for Contextualising hEterogeneous Sources} (SPACES), which includes the elements for representing sensors measurements that are independent from the sensors characteristics. Such set can reduce the effort for data fusion and interpretation, moreover it enforces both the reuse of existing infrastructure and the openness of the acquisition layer by providing a common framework for representing sensors readings.

\subsection{Acquisition Layer: TDSH}
TDSH underlying idea is that time should be a full-fledged first class object that applicative layers can observe and control.  Thus, TDSH includes abstractions able to: i) simplify the realisation of customisable data acquisition systems: ii) control and expose the acquisition rates at application level without exposing the underlaying complexity. 
The former point enables to model and develop large scale systems that can be easily tailored for the different contexts in which they are employed. The latter allows to face unexpected situations (such as emergencies) or to adapt to a change of context. For example, if a workout is detected the heart sampling rate should increase to better represent the current situation, as well as it should increase in case of supposed heart failure or arrhythmia and decrease in situations where the pulse is known to be lower (i.e., when sleeping).

\subsection{Normalization Layer: SPACES}
SPACES underlying idea is that applicative layers should not deal with intrinsics characteristics of acquisition devices, but should focus on the provided data, being able to understand and contextualise them without any information about the sensors that provided them. Moreover, the modification or introduction of new sources for similar information should be completely transparent to the existing applications. 

It follows that, once the data has been acquired, it should be represented in an homogeneous form regardless of the kind of the sensed information. 
At that aim, SPACES relies on the concept of space both to physically contextualize information and to represent its payload. This way, information is modeled by a couple of values: its position in a physical space (e.g., a triplet x, y, and z in a Cartesian space) and its payload in a suitable values space (e.g., a triplet Red, Green, and Blue in a color RGB space). Information is also contextualized in a temporal space in order to include the time in which has been sensed. Finally, mapping functions allow different applications to reason on their subjective spaces. For example, an application reasons in terms of Celsius space, whereas another in terms of Fahrenheit space; an application in terms of Cartesian space, another in terms of graph representation of a building. 

\section{Case study}
A TDSH implementation has been realized, tailored for embedded systems without the support of any real-time Operating System. Moreover, an hardware specific library has been implemented in order to run TDSH on a \textit{STM32F4-Discovery}, exploiting the HAL driver library to offer features based on the specific hardware, such as Analog to Digital converters and the Direct Memory Access data transmission mode. The SPACE architecture has been implemented as a Java library and includes an initial set of space models and mapping functions.

\begin{figure}[h]
\centering
\includegraphics[scale=0.35]{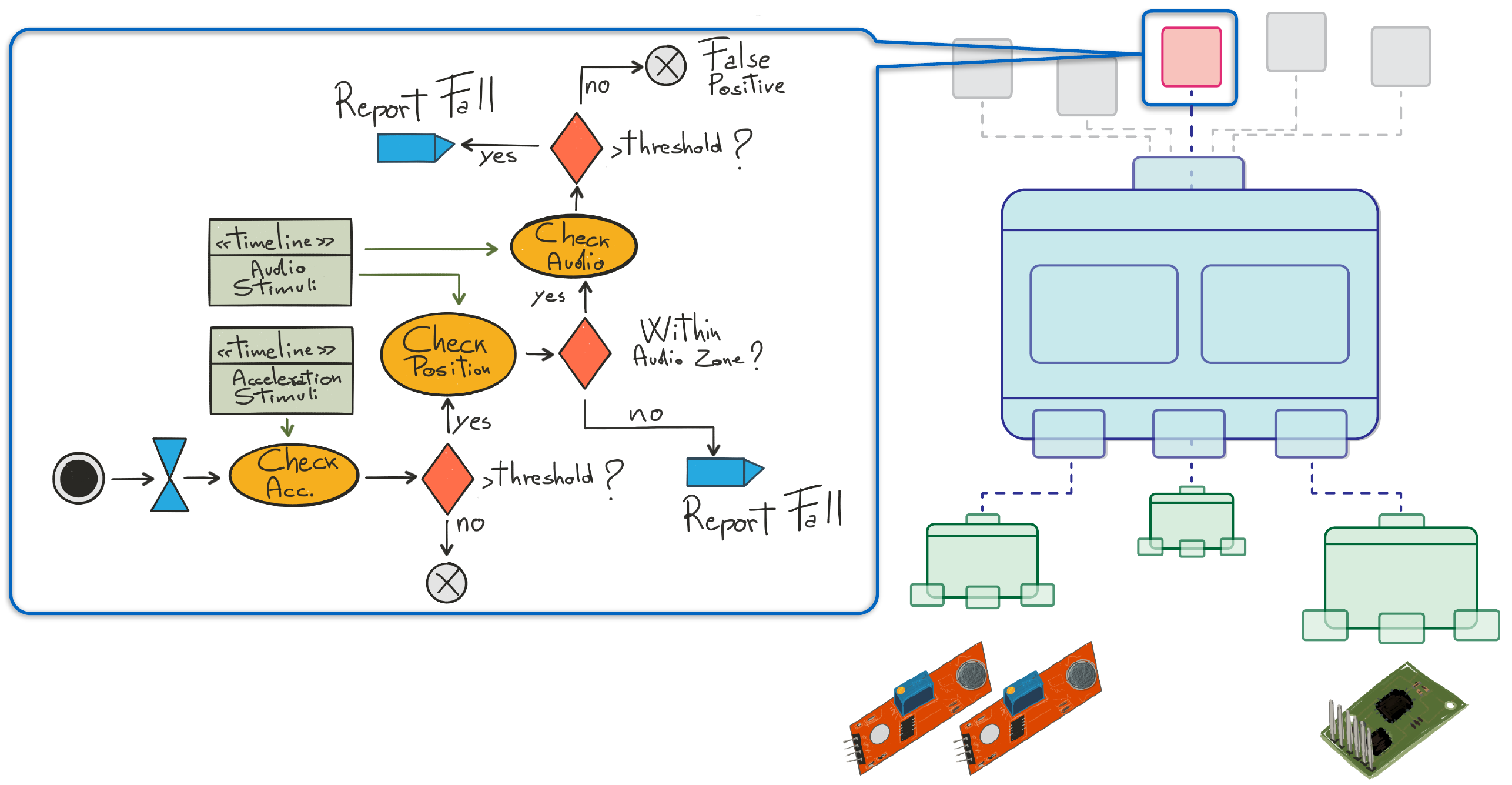}
\caption{An application scenario dealing with fall detection}
\label{fig:example}
\end{figure}

The architecture has been experimented in a concrete simplified scenario depicted in Figure~\ref{fig:example} and that includes an application able to recognize falls. Such an application has been chosen due to our knowledge of the domain~\cite{Micucci2017}. A wereable node hosts an accelerometer sensor and a microphonic array is installed in the environment.  Acoustic information from the array can be used to estimate the height of the source of a specific sound. A loud sound that is originated near the floor is more likely to be related to something or someone that is falling; a lound sound that has its origin in mid air may be something different such as music from a speaker.

\section{Conclusions}
This paper proposed two sets of architectural abstractions (TDSH and SPACES) that deal with acquisition of data from the field and that can be exploited to realize AAL systems. 


The initial test case demonstrates how TANA enables AAL applications to perform spatio-temporal reasoning.

We are implementing TDSH for the Android environment in order to use the sensors embedded smartphones for both Activity of Daily Living and falls recognition. Moreover, we are implementing the normalization activity as a set of RESTfull-based Web services.

\bibliographystyle{IEEEtran}
\bibliography{Bibliography}

\begin{thebibliography}{1}
\providecommand{\url}[1]{#1}
\csname url@samestyle\endcsname
\providecommand{\newblock}{\relax}
\providecommand{\bibinfo}[2]{#2}
\providecommand{\BIBentrySTDinterwordspacing}{\spaceskip=0pt\relax}
\providecommand{\BIBentryALTinterwordstretchfactor}{4}
\providecommand{\BIBentryALTinterwordspacing}{\spaceskip=\fontdimen2\font plus
\BIBentryALTinterwordstretchfactor\fontdimen3\font minus
  \fontdimen4\font\relax}
\providecommand{\BIBforeignlanguage}[2]{{%
\expandafter\ifx\csname l@#1\endcsname\relax
\typeout{** WARNING: IEEEtran.bst: No hyphenation pattern has been}%
\typeout{** loaded for the language `#1'. Using the pattern for}%
\typeout{** the default language instead.}%
\else
\language=\csname l@#1\endcsname
\fi
#2}}
\providecommand{\BIBdecl}{\relax}
\BIBdecl

\bibitem{un_world_2013}
{UN}, ``World {{Population Ageing}} 2013,'' United Nations, Tech. Rep.

\bibitem{kleinberger_ambient_2007}
T.~Kleinberger, M.~Becker, E.~Ras, A.~Holzinger, and P.~M{\"u}ller, ``Ambient
  intelligence in assisted living: Enable elderly people to handle future
  interfaces,'' in \emph{Universal Access in Human-Computer Interaction.
  {{Ambient}} Interaction}.\hskip 1em plus 0.5em minus 0.4em\relax {Springer},
  2007.

\bibitem{calvaresi_exploring_2016}
D.~Calvaresi, D.~Cesarini, P.~Sernani, M.~Marinoni, A.~F. Dragoni, and
  A.~Sturm, ``Exploring the ambient assisted living domain: a systematic
  review,'' \emph{Journal of Ambient Intelligence and Humanized Computing},
  vol.~8, no.~2, pp. 239--257, 2017.

\bibitem{Micucci2017}
D.~Micucci, M.~Mobilio, P.~Napoletano, and F.~Tisato, ``Falls as anomalies? an
  experimental evaluation using smartphone accelerometer data,'' \emph{Journal
  of Ambient Intelligence and Humanized Computing}, vol.~8, no.~1, pp. 87--99,
  2017.

\end{thebibliography}

\end{document}